\documentclass[preprint,prd,aps,superscriptaddress,11pt]{revtex4-1}

\usepackage{color}
\usepackage{graphicx}
\usepackage{amsmath}
\usepackage{amssymb}
\usepackage{dcolumn}
\usepackage{bm}

\newcommand{\be}{\begin{equation}}
\newcommand{\ee}{\end{equation}}
\newcommand{\ba}{\begin{eqnarray}}
\newcommand{\ea}{\end{eqnarray}}
\newcommand{\no}{\nonumber \\}
\newcommand{\gsim}{\mathrel{\hbox{\rlap{\lower.55ex \hbox {$\sim$}}
                   \kern-.3em \raise.4ex \hbox{$>$}}}}
\newcommand{\lsim}{\mathrel{\hbox{\rlap{\lower.55ex \hbox {$\sim$}}
                   \kern-.3em \raise.4ex \hbox{$<$}}}}

\def\vE{{\vec E}}

\def\roughly#1{\mathrel{\raise.3ex\hbox{$#1$\kern-.75em%
\lower1ex\hbox{$\sim$}}}}
\def\lsim{\roughly<}
\def\gsim{\roughly>}
\def\fm{{\text{fm}}}
\def\MeV{{\text{MeV}}}

\def\({\left(}
\def\){\right)}
\def\[{\left[}
\def\]{\right]}

\def\d{{\delta}}

\def\o{{\omega}}

\def\e{{\epsilon}}

\def\m{{\mu}}

\def\s{{\sigma}}

\def\t{{\tau}}

\def\x{{\xi}}

\newcommand{\pd}{{\partial}}
\newcommand{\pr}{\parallel}

\date{\today}

\begin{document}

\title{\bf On Conductivities of Magnetic Quark-Gluon Plasma at Strong Coupling}

\author{Wei Li}
\email{weili@stu2014.jnu.edu.cn}
\affiliation{Siyuan Laboratory, Physics Department, Jinan University, Guangzhou 510632, China}
\author{Shu Lin}
\email{linshu8@mail.sysu.edu.cn}
\affiliation{School of Physics and Astronomy, Sun Yat-Sen University, Zhuhai 519082, China}
\author{Jiajie Mei}
\email{jiajiemei@outlook.com}
\affiliation{School of Physics and Astronomy, Sun Yat-Sen University, Zhuhai 519082, China}


\begin{abstract}
  In the presence of strong magnetic field, the quark gluon plasma is magnetized, leading to anisotropic transport coefficients. In this work, we focus on effect of magnetization on electric conductivity, ignoring possible contribution from axial anomaly. We generalize longitudinal and transverse conductivities to finite frequencies. For transverse conductivity, a separation of contribution from fluid velocity is needed. We study the dependence of the conductivities on magnetic field and frequency using holographic magnetic brane model. The longitudinal conductivity scales roughly linearly in magnetic field while the transverse conductivity is rather insensitive to magnetic field. Furthermore, we find the conductivities can be significantly enhanced at large frequency. This can possibly extend lifetime of magnetic field, which is a key component of chiral magnetic effect.
\end{abstract}

\maketitle


\newpage

\section{Introduction}

Relativistic hydrodynamics has been remarkably useful in describing bulk evolution of quark-gluon plasma produced in heavy ion collisions. Since the early success of ideal hydrodynamics in describing elliptic flow \cite{Teaney:2001av,Kolb:2003dz}, there have been continuous efforts in formulate a hydrodynamics with higher accuracy and wider regime of applicability. Both kinetic theory approach and holographic approach have been used, which lead to significant development of the framework of relativistic hydrodynamics over the past decade. These include transient hydrodynamics \cite{Denicol:2012cn,Noronha:2011fi,Denicol:2016bjh,Denicol:2017lxn}, anisotropic hydrodynamics \cite{Alqahtani:2017mhy}, resummed hydrodynamics \cite{Lublinsky:2009kv,Heller:2013fn}, hydrodynamics with critical modes \cite{Stephanov:2017ghc,Attems:2018gou}, see \cite{Florkowski:2017olj,Romatschke:2009im} for a comprehensive review.

Recently, it has been realized that a strong magnetic field can be produced in off-central heavy ion collisions. The magnetic field plays an important role in the description of anomalous transport phenomena, in particular the chiral magnetic effect \cite{Kharzeev:2007jp,Son:2009tf,Neiman:2010zi}.
There have been growing efforts in applying hydrodynamics to study chiral magnetic effect \cite{Hirono:2014oda,Jiang:2016wve,Shi:2017cpu,Lin:2018nxj}. These studies assume a weak magnetic field such that the system remains isotropic. For strong magnetic field, both pressure and transports become anisotropic.
A systematic modification of the current hydrodynamics framework to the so called magnetohydrodynamics (MHD) is needed. This has been carried out by Hernandez and Kovtun (HK)\cite{Hernandez:2017mch}, see also dual formulation \cite{Grozdanov:2016tdf} and early works \cite{Huang:2009ue,Huang:2011dc,Critelli:2014kra,Finazzo:2016mhm}. The MHD including effect of axial anomaly is constructed in \cite{Hattori:2017usa}. Evaluation of anisotropic transport coefficients is needed for application of MHD. Viscosities in magnetic quark gluon plasma have been studied in \cite{Li:2017tgi,Hattori:2017qih}. Another interesting transport coefficient is the electric conductivity. In the presence of magnetic field, it splits into longitudinal and transverse conductivities. The longitudinal conductivity has been calculated at weak coupling by lowest Landau approximation in \cite{Hattori:2016lqx,Hattori:2016cnt} and beyond lowest Landau approximation in \cite{Fukushima:2017lvb}, see also conductivity from a quasi-particle model based on lowest Landau approximation \cite{Kurian:2017yxj}. At strong coupling, the longitudinal conductivity has been calculated in \cite{Arciniega:2013dqa,Patino:2012py,Jahnke:2013rca}. The conductivity in $2+1$ dimensional plasma has been obtained in \cite{Hartnoll:2007ip}. The isotropic conductivity in deconfined phase has also been calculated by lattice simulation \cite{Aarts:2007wj,Ding:2010ga,Buividovich:2010tn,Amato:2013naa,Aarts:2014nba,Ding:2016hua}.

The situation of transverse conductivity is quite different. The corresponding Kubo formula for longitudinal and transverse conductivities are derived by HK \cite{Hernandez:2017mch}, assuming B field in the $y$-direction and charge neutrality of plasma\footnote{The Kubo formulas are for fluid under external magnetic field. This is the set suitable for our holographic model study.}:
\begin{align}\label{kubo_HK}
&\lim_{\o\to0}\frac{1}{\o}\text{Im}G_{yy}(\o,{\vec k}=0)=\s_{\parallel}, \no
&\lim_{\o\to0}\frac{1}{\o}\text{Im}G_{xx}(\o,{\vec k}=0)=\o^2\frac{w_0^2}{B^4\s_\perp},
\end{align}
where $w_0=\e+p_\pr$ is the enthalpy density in equilibrium, $\s_\parallel$ and $\s_\perp$ are longitudinal and transverse conductivities respectively. The appearance of $\s_\perp$ in the denominator may seem odd. Essentially this is due to the interplay between transverse current and fluid velocity. It holds in the regime $\o\ll T$ and $\o\ll B/T$. The former is the hydrodynamic limit while the latter requires the B field to be not too small. $B/T$ can be regarded as inverse of time scale for cyclotron motion of plasma particles.

The aim of this work is to calculate both longitudinal and transverse conductivities in holographic magnetic quark-gluon plasma model. The paper is organized as follows: In Section II, we give an intuitive derivation of the Kubo formula for both transverse and longitudinal conductivities. The derivation naturally generalize conductivities in the hydrodynamic limit to finite frequency regime. Section III is devoted to the calculation of conductivities in holographic magnetic brane model. We discuss our results and phenomenological implications in Section IV.

\section{Kubo formulas}

We can reproduce the transverse Kubo formula in the following intuitive way: let us turn on a weak and slow varying homogeneous electric field $E$ along $x_1$-direction. The positive and negative charged particles will move in $\pm x_2$ direction. By the Lorentz force in the B field, both positive and negative particles gain momentum along $x_2$. This induces a net flow along $x_2$. No net flow is generated along $x_1$ due to the neutrality of plasma. The net effect of the flow along $x_2$ will cancel the current along $x_1$, again due to Lorentz force. This is the reason why transverse conductivity enters current only at higher order in $\o$.

We can formulate it more rigorously in the homogeneous limit
\begin{align}\label{const}
&j_i=\(E_i+\e_{ijk}v_jB_k\)\s_\perp+\pd_tP_i, \no
&T^{0i}=(\e+p_\pr)v_i-\e_{ijk}E_jM_k, \no
&\pd_tT^{0i}=\e_{ijk}j_jB_k.
\end{align}
Here the current $j_i$ consists of conducting current and polarization current, with $E_i+\e_{ijk}v_jB_k$ being effective field experienced by plasma particles and $P_i$ being electric polarization vector. The energy flow $T^{0i}$ contains fluid comoving contribution and medium contribution to Poynting vector, with $M_i$ being magnetization vector. The third equation is momentum non-conservation equation due to Lorentz force.
When medium in equilibrium has magnetization only, electric polarization is only induced by motion of fluid \cite{degroot,Caldarelli:2008ze}:
\begin{align}
P_i=\e_{ijk}v_jM_k,
\end{align}
as is required by Lorentz symmetry. To compare with HK, we note $M_\m=2p_{,B^2}B_\m$ and $p_\perp=p-MB$. \eqref{const} reproduces the constitutive equations of HK \cite{Hernandez:2017mch}. \eqref{const} is slightly more general in the sense that $v_i$ and $\s_\perp$ can be $\o$-dependent, thus \eqref{const} in fact defines transverse conductivity at finite frequency. Note that the use of fluid velocity at finite frequency is in the same spirit of resummed hydrodynamics \cite{Lublinsky:2009kv,Bu:2014sia,Bu:2014ena,Bu:2015ame}.

We can then solve for $v_i$:
\begin{align}
v_i=\frac{\e_{ijk}E_j\(B_k\s_\perp-i\o M_k\)}{B^2\s_\perp-i\o(\e+p_\pr+MB)}.
\end{align}
This gives the following current
\begin{align}
j_i=\frac{\o E_i\((\e+p_\pr- M\cdot B)\s_\perp+ i\o M^2\)}{iB^2\s_\perp+\o(\e+p_\pr+MB)}.
\end{align}
Note that $\vE=\pd_t{\vec A}=-i\o{\vec A}$. We readily obtain the correlator for transverse current:
\begin{align}\label{kubo_omega}
G_{xx}=-\frac{\d J_x}{\d A_x}=\frac{i\o^2\((\e+p_\pr-M\cdot B)\s_\perp+i\o M^2\)}{iB^2\s_\perp+\o(\e+p_\pr+MB)}.
\end{align}
Expanding \eqref{kubo_omega} in $\o$, we easily obtain:
\begin{align}\label{omega_series}
G_{xx}=\frac{(\e+p_\pr-MB)\o^2}{B^2}+i\frac{(\e+p_\pr)^2}{B^4\s_\perp}\o^3+\cdots.
\end{align}
We immediately see \eqref{omega_series} gives Kubo formula for $\s_\perp$ in \eqref{kubo_HK}. However it is singular as $B\to0$ due to non-commutativity of hydrodynamic limit and isotropic limit. \eqref{kubo_omega} can be safely used in both limits.
We solve the $\o$ dependent conductivity as
\begin{align}\label{sigma_perp}
  \s_\perp(\o,B)=\frac{i\o\(\(\e+p_\pr+MB\)G_{xx}+\o^2M^2\)}{B^2G_{xx}-(\e+p_\pr-MB)\o^2}.
\end{align}

The case of longitudinal conductivity is trivial because Lorentz force is not relevant. The corresponding Kubo formula is given by
\begin{align}\label{sigma_pr}
  \s_\pr(\o,B)=\frac{G_{yy}}{i\o}.
\end{align}

\section{The holographic computation of conductivities}

\subsection{Magnetic brane background}
We use magnetic brane background \cite{DHoker:2009mmn} for the computation of conductivities. The background is a solution to five-dimensional Einstein-Maxwell theory with a negative cosmological constant\footnote{Note that the normalization of $U(1)$ field differs by a factor of $4$ from the standard electromagnetic field. We will stick to this normalization, which does not alter our results}:
\begin{equation}
S=\frac{1}{16\pi G_5}\Bigg[\int\mathrm{d}^5x\sqrt{-g}\Big(R+\frac{12}{L^2}-L^2F_{MN}F^{MN}\Big)-k\int A\wedge F\wedge F\Bigg].
\end{equation}
Here $L$ is the $AdS$ radius set to unity below, $F=\mathrm{d}A$ is the Maxwell field strength, and the second term in the action corresponds to Chern-Simons term. The Chern-Simons term corresponds to axial anomaly. The axial anomaly is known to lead to negative magnetoresistance \cite{Son:2012bg,Landsteiner:2014vua}. In this study, we wish to focus on contribution from magnetization. To this end, we turn off the Chern-Simons term. The resulting equations of motion (EOM) read
\begin{align}\label{eom}
&R_{MN}+4g_{MN}+\frac{1}{3}F^{PQ}F_{PQ}g_{MN}-2F_{MP}F_{N}^{~P}=0,\notag\\
&\nabla_MF^{MN}=0.
\end{align}
The magnetic solution is given by \cite{DHoker:2009mmn}
\begin{align}\label{rcoord}
\mathrm{d}s^2&=-U(r)\mathrm{d}t^2+\frac{\mathrm{d}r^2}{U(r)}+e^{2V(r)}\Big((\mathrm{d}x^1)^2+(\mathrm{d}x^2)^2\Big)+e^{2W(r)}\mathrm{d}y^2,\notag\\
F&=B\mathrm{d}x^1\wedge\mathrm{d}x^2.
\end{align}
The warping factor $U(r)$ contains a zero at $r=r_H$, which is the location of horizon. This corresponds to a temperature of the plasma $T_H=\frac{U'(r_H)}{4\pi}$. The solution of the background can only be obtained numerically. It is convenient to compactify the radial coordinate by defining $r=r_H u^{-1/2}$, which puts the horizon at $u=1$. The background in terms of $u$ coordinate becomes
\begin{align}
\mathrm{d}s^2&=-U(u)\mathrm{d}t^2+\frac{\mathrm{d}u^2}{4u^3U(u)}+e^{2V(u)}(\mathrm{d}(x^1)^2+\mathrm{d}(x^2)^2)+e^{2W(u)}\mathrm{d}y^2,\notag\\
F&=B\mathrm{d}x^1\wedge\mathrm{d}x^2.
\end{align}
The EOM read
\begin{align}\label{eom_magnetic}
&-2 B^2 e^{-4 V}+6 u^3
   U''+3 u^2 U' \left(4
   u V'+2 u
   W'+3\right)-12=0,\notag\\
&-\frac{4}{3} e^{-2 V}
   \left(B^2+3 u^3 e^{4 V}
   U' V'-3 e^{4
   V}\right)-2 u^2 U
   e^{2 V} \left(2 u
   V''+V' \left(2 u
   W'+3\right)+4 u
   V'^2\right)=0,\notag\\
&B^2-6 u^3 e^{4 V} U'
   W'-3 u^2 U e^{4 V}
   \left(\left(4 u
   V'+3\right) W'+2 u
   W''+2 u W'{}^2\right)+6
   e^{4 V}=0,
\end{align}
with the derivatives taken with respect to $u$.
The numerical solution is to be obtained by integrating the following horizon solution to the boundary of AdS:
\begin{align}\label{horizon_sol}
U(u)&= u_1(u-1)+u_2(u-1)^2+\cdots,\notag\\
V(u)&= v_0+v_1(u-1)+\cdots,\notag\\
W(u)&= w_0+w_1(u-1)+\cdots.
\end{align}
We can put $v_0=w_0=0$ by rescaling of $x$ and $y$ coordinates. We also put $u_1=-2$, which sets the unit by fixing the temperature to $T=\frac{1}{4\pi}$. The magnetic field after the rescaling is denoted as $b$, which is to replace $B$ in \eqref{eom_magnetic}. The higher order coefficients in \eqref{horizon_sol} can be determined recursively from EOM as:
\begin{align}
v_1=\frac{2}{3}(b^2-3)\qquad w_1=-\frac{1}{3}(b^2+6),\qquad u_2=\frac{1}{24}(10b^2-3).
\end{align}
For a particular $b$ we can numerically solve the metric functions. Near boundary, the solution behaves like
\begin{equation}
U\sim\frac{1}{u},\qquad e^{2V}\sim\frac{v(b)}{u},\qquad e^{2W}\sim\frac{w(b)}{u}
\end{equation}
Thus we need the following rescaling $t\to\hat{t}=t$, $x_1\to\hat{x}_1=\sqrt{v(b)}x_1$, $x_2\to\hat{x}_2=\sqrt{v(b)}x_2$, $y\to\hat{y}=\sqrt{w(b)}y$ to bring the background to the standard AdS asymptotics. After the rescaling, the full background reads
\begin{align}
\mathrm{d}s^2&=-\tilde{U}(u)\mathrm{d}\hat{t}^2+\frac{\mathrm{d}u^2}{4u^3\tilde{U}(u)}+e^{2\tilde{V}(u)}(\mathrm{d}\hat{x}_1^2+\mathrm{d}\hat{x}_2^2)+e^{2\tilde{W}(u)}\mathrm{d}\hat{y}^2,\notag\\
F&=B\mathrm{d}\hat{x}_1\wedge\mathrm{d}\hat{x}_2,
\end{align}
where 
\begin{equation}
B=\frac{b}{v(b)},\quad e^{2\tilde{V}(u)}=\frac{e^{2V(u)}}{v(b)},\quad e^{2\tilde{W}(u)}=\frac{e^{2W(u)}}{w(b)}.
\end{equation}
Below we use tilded symbols for metric functions with standard AdS asymptotics.

\subsection{Transverse and longitudinal conductivities}
To calculate transverse conductivity, we consider the following linear perturbation about the background
\begin{align}
\delta g_{tx_2}&=h_{tx_2}(u)e^{-i\omega t},\notag\\
\delta A_{x_1}&=a_{x_1}(u)e^{-i\omega t}.
\end{align}
It is convenient to use metric perturbation with mixed indices $h_t^{x_2}=e^{-2\tilde{V}(u)}h_{tx_2}(u)\equiv G_{tx_2}(u)$. After substituting into the equations of motion we obtain the following ordinary differential equations
\begin{align}\label{peom}
&a_{x_1}^{''}(u)+\frac{a_{x_1}^{'}(u)(4u^3\tilde{U}(u)\tilde{U}^{'}(u)+4u^3\tilde{U}(u)^2\tilde{W}^{'}(u)+6u^2\tilde{U}(u)^2)}{4u^3\tilde{U}(u)^2}+\frac{\omega^2a_{x_1}(u)}{4u^3\tilde{U}(u)}+\frac{iB\omega G_{tx_2}(u)}{4u^3\tilde{U}(u)^2}=0,\notag\\
&G_{tx_2}^{''}(u)+\frac{G_{tx_2}^{'}(u)(8u\tilde{V}^{'}(u)+2u\tilde{W}^{'}(u)+3)}{2u}-\frac{G_{tx_2}(u)e^{-4\tilde{V}(u)}B^2}{u^3\tilde{U}(u)}+\frac{iB\omega a_{x_1}(u)e^{-4\tilde{V}(u)}}{u^3\tilde{U}(u)}=0,\notag\\
&4B\tilde{U}(u)a_{x_1}^{'}(u)+i\omega e^{4\tilde{V}(u)}G_{tx_2}^{'}(u)=0.
\end{align}
Near the horizon, the solution behave as $a_{x_1}\sim(1-u)^\alpha,G_{tx_2}\sim(1-u)^\beta$. The incoming exponent is given by $\alpha=-i\omega,\beta=\alpha+1$. We will look for solution of the form
\begin{align}\label{sol1}
a_{x_1}^{inc}(u)&=\Big(1-u^2\Big)^{\alpha}A(u),\notag\\
G_{tx_2}^{inc}(u)&=\Big(1-u^2\Big)^{1+\alpha}G(u).
\end{align}
Near the boundary, the incoming wave solution behaves like
\begin{align}
a_{x_1}^{inc}(u)&\sim A(u)\sim A^{(0)}+A^{(1)}u-\frac{1}{4}\Big(A^{(0)}\omega^2+iBG^{(0)}\omega\Big)u\log(u)+\cdots,\notag\\
G_{tx_2}^{inc}(u)&\sim G(u)\sim G^{(0)}+G^{(2)}u^2+\Big(2B^2G^{(0)}-iA^{(0)}B\omega\Big)u^2\log(u)+\cdots.
\end{align}
In fact this set of incoming solution is determined by only one parameter, which does not match the number of unknown fields. In fact, we can find another constant solution
\begin{equation}\label{sol2}
a_{x_1}^{con}(u)=C_2,\qquad G_{tx_2}^{con}(u)=\frac{i\omega C_2}{B}.
\end{equation}
This is a pure gauge solution of the following type
\begin{align}
  a_M=\x^N\pd_NA_M+\pd_M\x^NA_N,\qquad h_N^M=\nabla_N\x^M+\nabla^M\x_N.
\end{align}
Fixing the background gauge field as $A_M=-Bx_2\d_M^1$, we find the constant solution \eqref{sol2} is given by $\x^M=\d^M_2e^{-i \o t}$. Note that the gauge choice of the background is necessary to ensure the vanishing of all other perturbations.
Thus the general solution is a linear combination of these two solutions.
\begin{align}
a_{x_1}(u)&=a_{x_1}^{inc}(u)+a_{x_1}^{con}(u),\notag\\
G_{tx_2}(u)&=G_{tx_2}^{inc}(u)+G_{tx_2}^{con}(u).
\end{align}
In order to calculate the retarded correlator $G^R_{x_1x_1}$, we need to eliminate the contribution to current from response to metric perturbation, thus we should turn off boundary value of metric perturbation. It amounts to setting $\lim_{u\to0}G_{tx_2}(u)=0$. This fixes $C_2$ to
\begin{equation}
C_2=\frac{iBG^{(0)}}{\omega}.
\end{equation}
Therefore the retarded correlator $G^R_{x_1x_1}$ reads
\begin{equation}\label{gxx}
G^R_{x_1x_1}(\omega,\vec{k}=0)=\frac{1}{2\pi G_5}\left(\frac{\omega A^{(1)}}{\omega A^{(0)}+iBG^{(0)}}\right).
\end{equation}

To calculate the longitudinal conductivity (in $y$ direction), we only have to consider the following perturbation
\begin{equation}
\delta A_{y}=a_y(u)e^{-i\omega t}.
\end{equation}
The perturbed field $a_y(u)$ satisfies the following EOM
\begin{equation}
a_y^{''}(u)+a_y^{'}(u)\left(\frac{\tilde{U}^{'}(u)}{\tilde{U}(u)}+2\tilde{V}^{'}(u)-\tilde{W}^{'}(u)+\frac{3}{2u}\right)+\frac{\omega^2a_y(u)}{4u^3\tilde{U}(u)^2}=0.
\end{equation}
Near the horizon, the incoming wave solution behaves as $a_y\sim(1-u)^{-i\omega}$. We look for the solution of the form
\begin{equation}\label{B}
a_y(u)=(1-u^2)^{-i\omega}D(u).
\end{equation}
Near the boundary, the solution behaves like
\begin{equation}
a_y(u)=D^{(0)}+D^{(1)}u-\frac{D^{(0)}\omega^2}{4}u\log(u)+\cdots.
\end{equation}
Therefore the retarded correlator $G_{yy}$ reads
\begin{equation}\label{gyy}
G_{yy}^R(\omega,\vec{k}=0)=\frac{1}{2\pi G_5}\left(\frac{D^{(1)}}{D^{(0)}}\right).
\end{equation}
We will study \eqref{gxx} and \eqref{gyy} in different regimes in the following.

\subsection{Hydrodynamic regime}
In hydrodynamic regime we can solve the equation perturbatively in $\omega$,
\begin{align}
A(u)&=A_0(u)+i\omega A_1(u)+\omega^2A_2(u)+\cdots,\notag\\
G(u)&=G_0(u)+i\omega G_1(u)+\cdots,\notag\\
D(u)&=D_0(u)+i\omega D_1(u)+\cdots,
\end{align}
where $A,G,D$ are defined in (\ref{B}) and (\ref{sol1}).
Let us study transverse equations first. The coupled EOM of $A$ and $G$ read\begin{align}\label{eom_tran}
&A_0'(u)=0,\notag\\
&\frac{8BuA_0(u)\tilde{U}(u)}{u^2-1}-4B\tilde{U}(u)A_1'(u)+(u^2-1)e^{4\tilde{V}(u)}G_0'(u)+2uG_0(u)e^{4\tilde{V}(u)}=0,\notag\\
&\frac{8BA_1(u)\tilde{U}(u)}{u^2-1}+4B\tilde{U}(u)A_2'(u)-2uG_0(u)e^{4\tilde{V}(u)}+u^2e^{4\tilde{V}(u)}G_1'(u)+2uG_1(u)e^{4\tilde{V}(u)}=0,\notag\\
&\frac{2B^2(u^2-1)G_0(u)e^{-\tilde{V}(u)}}{u^3\tilde{U}(u)}-2(u^2-1)G_0''(u)-\frac{\left((u^2-1)G_0'(u)+2uG_0(u)\right)\left(8u\tilde{V}'(u)+2u\tilde{W}'(u)+3\right)}{u}\notag\\
&-8uG_0'(u)-4G_0(u)=0,\notag\\
&-\frac{1}{u}\left(-2uG_0(u)+(u^2-1)G_1'(u)+2uG_1(u)\right)(8u\tilde{V}'(u)+2u\tilde{W}'(u)+3)+\frac{2BA_0(u)e^{-\tilde{4}V(u)}}{u^3U(u)}\notag\\
&+\frac{2B^2(u^2-1)G_1(u)e^{-4\tilde{V}(u)}}{u^3\tilde{U}(u)}+8u\left(G_0'(u)-G_1'(u)\right)+\frac{8u^2G_0(u)}{u^2-1}-2(u^2-1)G_1''(u)=0.
\end{align}
We first expand the fields $h_{tx_2}$ and $a_{x_1}$ and background solution $\tilde{U}(u),\tilde{V}(u),\tilde{W}(u)$ near horizon using (\ref{horizon_sol}). And then we numerically solve (\ref{eom_tran}) by giving an initial condition on the horizon that $A_0(1)=1$. This fixes normalization of the solution but does not affect result of correlators. Note that $A_0(u)$ has a constant solution $A_0(u)=1$ with this specific initial condition. We further require all higher order functions vanish on the horizon.
The perturbative solution give the following perturbative expansion of transverse retarded correlator
\begin{align}\label{gxx_omega}
&G_{x_1x_1}^R(\omega,\vec{k}=0)=\omega^2\frac{A_1^{(1)}}{BG_0^{(0)}}+i\omega^3\frac{A_1^{(1)}BG_1^{(0)}-A_1^{(1)}+A_2^{(1)}BG_0^{(0)}}{B^2(G_0^{(0)})^2}+O(\omega^{4}).
\end{align}
Here the functions $A_i^{(0)}$, $A_i^{(1)}$, $G_i^{(0)}$ and $G_i^{(1)}$ are defined through the following boundary expansions
\begin{align}
  &A_i(u)=A_i(u)^{(0)}+uA_i(u)^{(1)}+\cdots,\quad i=0,1,2,\notag\\
  &G_i(u)=G_i(u)^{(0)}+u^2G_i(u)^{(1)}+\cdots,\quad i=0,1.
\end{align}
\eqref{gxx_omega} is the expected form of transverse correlator in hydrodynamic regime. The imaginary part starts from $\o^3$, whose coefficient can be used to determine transverse conductivity with the corresponding Kubo formula in \eqref{kubo_HK}.

\begin{figure}
\includegraphics[width=0.8\textwidth]{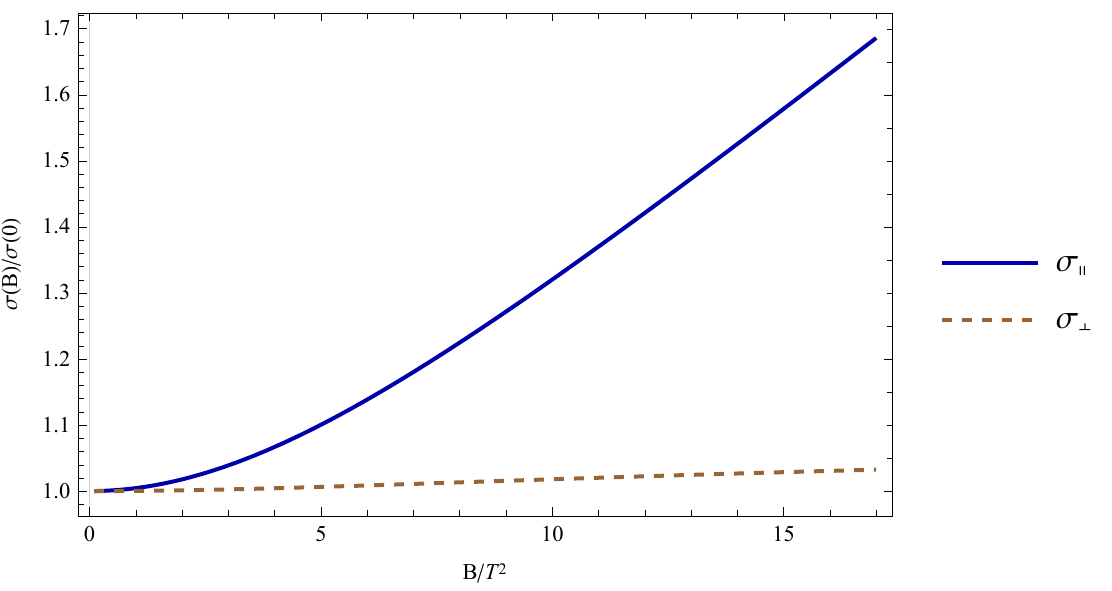}
\caption{\label{sigma_B}The dependence of $\s_\pr$ and $\s_\perp$ on $B$. At large $B$, $\s_\pr$ grows linearly with $B$, while $\s_\perp$ tends to a constant. At small $B$, $\s_\pr$ and $\s_\perp$ agree with each other, although strictly speaking we cannot take the limit $B\to0$ in hydrodynamic regime.}
\end{figure}

The longitudinal equation can be studied similarly. The EOM in terms of $D_0$ and $D_1$ are given by
\begin{align}\label{eom_long}
&2uD_0'(u)\tilde{U}'(u)+\tilde{U}(u)\left(2uD_0''(u)+D_0'(u)(4u\tilde{V}'(u)-2u\tilde{W}'(u)+3)\right)=0,\notag\\
&2u\left(D_1'(u)-\frac{2u}{u^2-1}\right)\tilde{U}'(u)+\tilde{U}(u)\Bigg(\left(2u(D_1''(u)+\frac{2(u^2+1)}{(u^2-1)^2}\right)\notag\\
&+\left(D_1'(u)-\frac{2u}{u^2-1}\right)\left(4u\tilde{V}'(u)-2u\tilde{W}'(u)+3\right)\Bigg)=0.
\end{align}
Again we numerically solve (\ref{eom_long}) by giving the initial condition that $D(1)=1$. For $D_0(u)$, we find that $D_0'(1)=D_0''(1)=0$, thus it admits a constant solution $D_0(u)=1$. Then we numerically solve $D_1(u)$.
The perturbative solution gives the following perturbative expansion of longitudinal retarded correlator
\begin{align}
&G_{yy}^R(\omega,\vec{k}=0)=i\omega D_1^{(1)}+O(\omega),
\end{align}
where $D_1^{(1)}$ is defined through boundary expansion of $D_1$
\begin{align}
  &D_1(u)=D_1^{(0)}+uD_1^{(1)}+\cdots.
\end{align}
Our boundary condition fixes $D_1^{(0)}=0$.
We can thus simply identify $D_1$ with longitudinal conductivity in the hydrodynamic regime based on \eqref{kubo_HK}.
We show in Fig.~\ref{sigma_B} the dependence of $\s_\pr$ and $\s_\perp$ on $B$. We observe nearly linear dependence of $\s_\pr$ on $B$. This is consistent with the picture that all the charge carriers are from the lowest Landau level in large $B$ limit, with the density of charge carriers proportional to $B$. On the other hand, $\s_\perp$ tends to a constant at large $B$. Although we cannot take the limit $B\to0$ in hydrodynamic regime, we do find at small $B$, $\s_\pr$ and $\s_\perp$ are numerically consistent with each other. The two limits are also obtained in \cite{Mamo:2013efa}, although in that case, the mixing of perturbation in transverse case was not taken into account.
Interestingly, the approach of \cite{Mamo:2013efa} turns out to give the correct answer in hydrodynamic regime. We show this by membrane paradigm in appendix.

\subsection{Conductivities at arbitrary frequency}

\begin{figure}
  \includegraphics[width=0.8\textwidth]{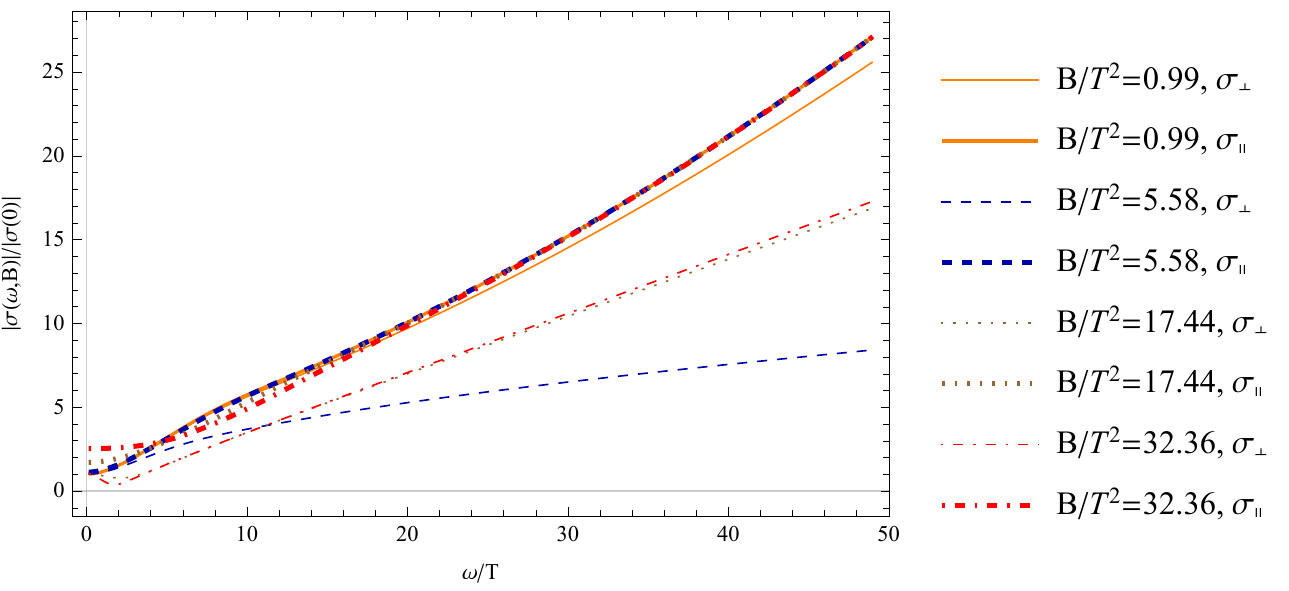}
\caption{\label{sigma_abs}The dependence of $|\s_\pr|$ and $|\s_\perp|$ on $\o$ for several $B$. At large $\o$, both $|\s_\pr|$ and $|\s_\perp|$ can be significantly larger than its hydrodynamic counterparts. The large $\o$ limits $|\s_\pr|$ and $|\s_\perp|$ show qualitative difference: The former has only weak dependence on $B$, and the latter depends on $B$ non-monotonically.}
\end{figure}

\begin{figure}
  \includegraphics[width=0.8\textwidth]{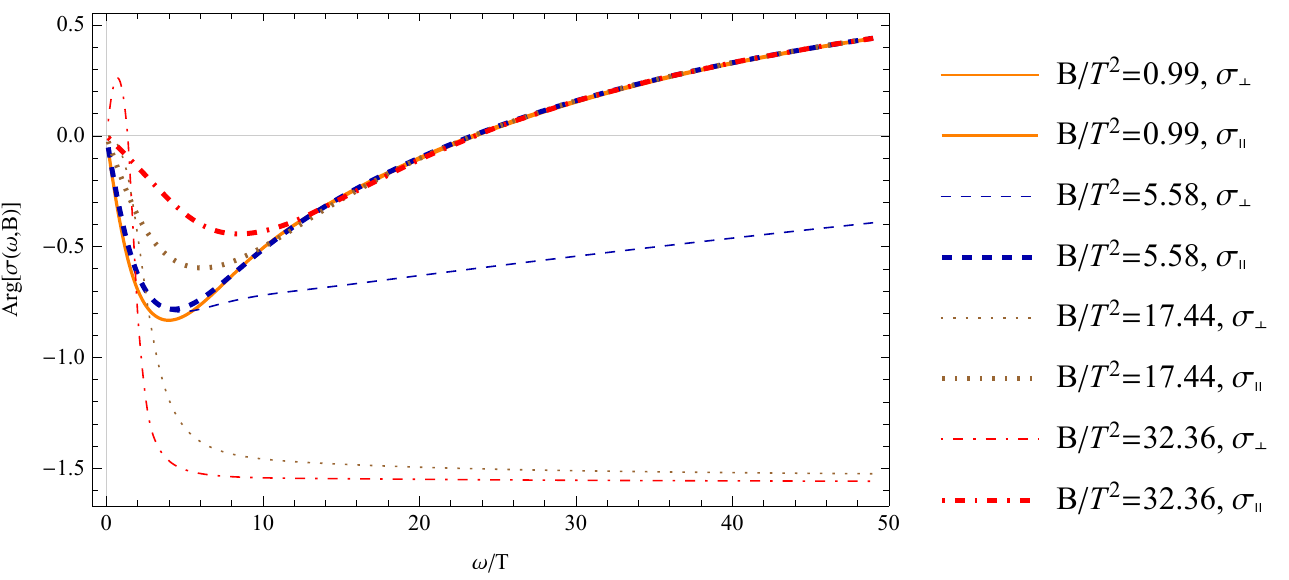}
\caption{\label{sigma_arg}The dependence of $\text{Arg}[\s_\pr]$ and $\text{Arg}[\s_\perp]$ on $\o$ for several $B$. The large $\o$ limit of $\s_\pr$ for different $B$ approach a universal curve. The large $\o$ limit of $\s_\perp$ has non-trivial dependence on $B$. At small $B$, it approaches the universal curve of $\s_\pr$; at intermediate $B$, the phase lags behind the universal curve; at large $B$, the phase lags approaches $-\pi/2$ numerically.}
\end{figure}

\begin{figure}
  \includegraphics[width=0.8\textwidth]{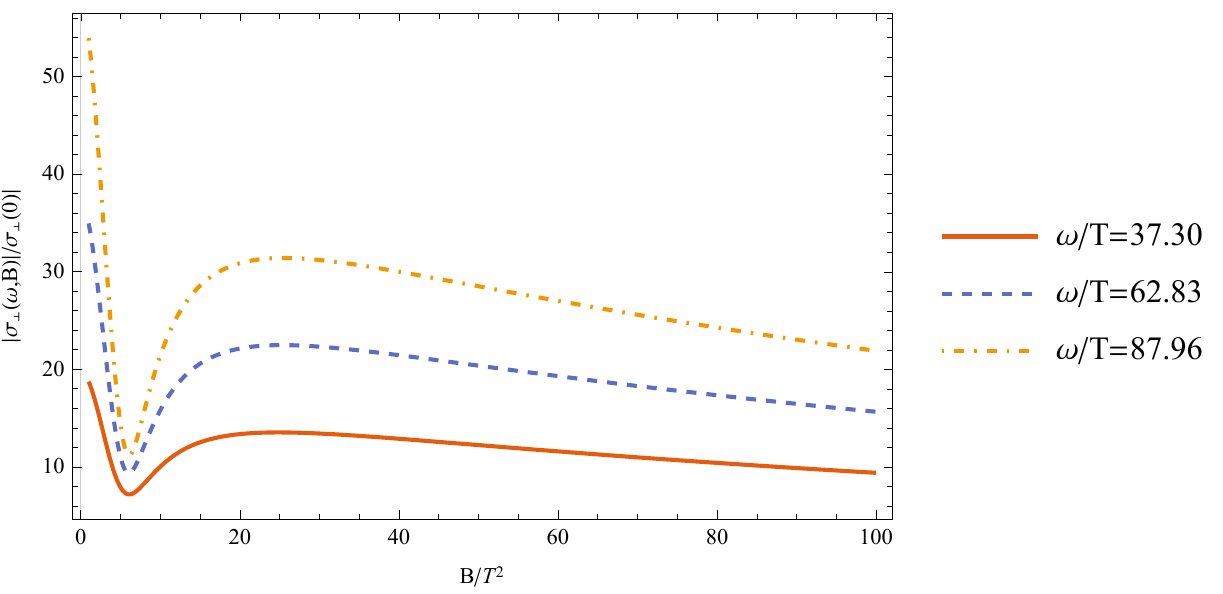}
  \caption{\label{B_perp}Non-monotonic $B$ dependence of $|\s_\perp|$ for three large values of $\o$. The minimum of $|\s_\perp|$ is seem to be independent of $\o$.}
\end{figure}

Beyond hydrodynamic regime, we should use \eqref{sigma_perp} and \eqref{sigma_pr} as definitions of conductivities at finite frequency. Note that beyond hydrodynamic regime, the conductivities are in general complex.
We solve the transverse and longitudinal EOM numerically to obtain complex conductivities. We plot $|\s|$ and $\text{Arg}[\s]$ as a function of $\o$ in Fig.~\ref{sigma_abs} and Fig.~\ref{sigma_arg}. $|\s|$ characterizes the magnitude of current induced in magnetic plasma by external electric field. Fig.~\ref{sigma_abs} shows both $|\s_\pr|$ and $|\s_\perp|$ can be significantly larger than their hydrodynamic counterparts at large $\o$. The large $\o$ limit of $|\s_\pr|$ is rather insensitive to $B$, while for $|\s_\perp|$, its large $\o$ limit is non-monotonic in $B$. We also plot the $B$-dependence of $|\s_\perp|$ at large values of $\o$ in Fig.~\ref{B_perp}. On the other hand, $\text{Arg}[\s]$ characterizes the phase difference of current and external electric field. As $\o\to0$, the conductivities are real meaning that the current is in phase with applied electric field. The large $\o$ limit of $\text{Arg}[\s_\pr]$ approaches a universal curve, independent of $B$. The large $\o$ limit of $\text{Arg}[\s_\perp]$ has non-trivial $B$ dependence: at small $B$, it approaches the same universal curve as $\text{Arg}[\s_\pr]$; at intermediate $B$, the phase of transverse current lags further behind; at large $B$, the phase lag approaches $-\pi/2$ numerically.

In fact, the large $\o$ limit of $\s_\pr$ can be obtained analytically by noting that $B\ll\o^2$ becomes irrelevant. Ignoring the magnetic field, we can use the known result for retarded current-current correlator (adapted to our choice of unit) \cite{Myers:2007we}:
\begin{align}
  G^R_{yy}=2i\o+4\o^2\(\Psi((1-i)\o)+\Psi(-(1+i)\o)\),
\end{align}
which gives us the following asymptotics of conductivity
\begin{align}
  \lim_{\o\to\infty}\s_\pr(\o)= 2i\(-i\pi-\ln(2)+2\ln(\o)\)\o.
\end{align}
$|\s_\pr|$ is linear in $\o$ up to logarithmic correction. $\text{Arg}[\s_\pr]$ approaches $\pi/2$ slowly from below. This is consistent with our numerical results in Fig.~\ref{sigma_abs} and Fig.~\ref{sigma_arg}.

The origin of the non-trivial $B$-dependence of $|\s_\perp|$ at large $\o$ is instructive. Note that for $B\to0$, $\s_\perp$ approach the same universal behavior as $\s_\pr$ at large $\o$. It is tempting to attribute the difference at finite $\o$ to the dynamics of magnetization to external electric field. Fig~\ref{sigma_abs} and Fig.~\ref{sigma_arg} seem to suggest the magnetization respond to longitudinal electric field weakly, but has non-trivial response to transverse electric field. It is also interesting to note that the minimum of $|\s_\perp|$ in Fig.~\ref{B_perp} corresponds to value of $B$ that maximizes the phase delay of the current in Fig.~\ref{sigma_arg}. More quantative studies are needed to understand the mechanism underlying this behavior.


\section{Discussion}

We study longitudinal and transverse conductivities at finite magnetic field $B$ and frequency $\o$. While the former is a straight forward generalization the static (hydrodynamic) limit, the latter involves a careful subtraction of fluid velocity contribution. We arrive at a Kubo formula that is applicable at finite frequencies. It reduces to Kubo formula in the hydrodynamic regime \cite{Hernandez:2017mch}. We focus on the effect of magnetization on conductivities ignoring possible contribution from axial anomaly.

Using holographic background dual to quark gluon plasma with external magnetic field, we study the $B$ and $\o$ dependence of conductivities. In the hydrodynamic regime, we find the longitudinal conductivity $\s_\pr$ scales linearly with $B$ at large $B$, consistent with lowest Landau level picture. The transverse conductivity is not sensitive to $B$ field in a wide region.
The $\o$ dependence of conductivities is more interesting. We find both conductivities scales nearly linearly in $\o$ at large $\o$. This could be understood qualitatively as the relaxation time increases with frequency of electric field. The $B$ dependence of the large $\o$ limits of $\s_\pr$ and $\s_\perp$ differ: The former is nearly independent of $B$, while the latter shows a non-monotonic dependence on $B$.

The obtained values of conductivities might be relevant for the physics of chiral magnetic effect \cite{Kharzeev:2007jp}. The effect of conductivity on lifetime of magnetic field is studied in \cite{McLerran:2013hla}. It is found that only very large conductivities can extend the lifetime of magnetic field. In heavy ion collisions experiment, the produced magnetic field \cite{Skokov:2009qp,Deng:2012pc} can be estimated as $B/T^2\simeq m_\pi^2/T^2\simeq 0.26$, with $T=350\MeV$. The magnetic field itself might not have significant effect on conductivity from Fig.~\ref{sigma_B}. However, the rapid decaying magnetic field induces rapid changing electric field, which calls for use of conductivities at finite frequency. Assuming a lifetime of magnetic field as $\t\simeq 1\fm$, we would obtain $\o/2\pi T\simeq 1/\t T\simeq 0.57$. At this frequency, the conductivities are enhanced by a factor of $3$ from Fig.~\ref{sigma_abs}. A lifetime of magnetic field $\t\simeq 0.2fm$ would lead to a factor of $10$ for the conductivity! A re-evaluation of the effect based on finite frequency conductivities is needed. We leave it for future analysis.

\section*{Acknowledgments}

S.L. is grateful to Yan Liu for useful discussions. J.J.M thanks Sun Yat-Sen University for hospitality during his visit when this work is completed. S.L. is supported by One Thousand Talent Program for Young Scholars and NSFC under Grant Nos 11675274 and 11735007.

\appendix

\section{Transverse conductivity from heat current correlator}

In the appendix, we obtain the conductivities using membrane paradigm \cite{Iqbal:2008by}. While conventional membrane paradigm works for $\s_\pr$, it fails for $\s_\perp$ due to mixing of current and energy flow. The resolution is that $\s_\perp$ can also be obtained from correlator of heat current. The corresponding Kubo formula is given by \cite{Hernandez:2017mch}
\begin{align}\label{kubo_E}
  \frac{1}{\o}ImG_{T_{0x}T_{0x}}=\frac{w_0^2}{\s_\perp B^2}.
\end{align}
For convenience we revert to the original $r$ coordinate (\ref{rcoord}).

To study current and energy flow in response to external electric field and metric perturbation, we turn on the following perturbations \cite{Donos:2014cya,Blake:2015ina,Avila:2018sqf}.
\begin{align}
A_{x_1}&=-Et+\delta a_{x_1}(r)\notag,\\
g_{tx_2}&=-\zeta tU(r)+\delta g_{tx2}(r)\notag,\\
g_{rx_2}&=\delta g_{rx_2}(r).
\end{align}
We can construct the heat current in the linear order following the procedure in \cite{Liu:2017kml},
\begin{equation}
\mathcal{Q}=2\sqrt{-g}G^{rx_2},
\end{equation}
Express it with perturbation fields we have
\begin{equation}\label{heat}
\mathcal{Q}=2\sqrt{-g}G^{rx_2}=U(r)^2e^{2V(r)-W(r)}\partial_r\left(\frac{\delta g_{tx2}(r)}{U(r)}\right).
\end{equation}
By the incoming wave condition and regularity on the horizon, the perturbation behaves like
\begin{align}
\delta a_{x1}(r)&=-\frac{E}{4\pi T}\log(r-r_h)+O(r-r_h),\notag\\
\delta g_{tx2}(r)&=U(r)\delta g_{rx2}(r)-\frac{\zeta U(r)}{4\pi T}\log(r-r_h)+O(r-r_h).
\end{align}
And we can solve $g_{rx2}$ use the Einstein equation
\begin{equation}
\delta g_{rx2}=-\zeta\frac{e^{6V(r)}}{4B^2U(r)}\partial_r(U(r)e^{-2V(r)})-\frac{e^{2V(r)}}{B}\delta a'_{x1}.
\end{equation}
Because the heat current $\mathcal{Q}$ satisfies $\partial_r\mathcal{Q}=0$, it can be evaluated at any location of $r$. Thus we evaluate it at the horizon $r_h$,
\begin{equation}
\mathcal{Q}=-E\frac{\pi T}{B}e^{W(r_h)+2V(r_h)}+\zeta\frac{\pi^2T^2}{B^2}e^{4V(r_h)+W(r_h)}.
\end{equation}
In neutral plasma where $\mu=0$, the Kubo formula reads
\begin{equation}
\lim_{\omega\to0}\frac{1}{\omega}ImG_{Q_{x}Q_{x}}^R(\omega,\vec{k}=0)=\lim_{\omega\to0}\frac{1}{\omega}ImG_{T_{0x}T_{0x}}^R(\omega,\vec{k}=0)=\frac{1}{4\pi G_5}\frac{\partial\mathcal{Q}}{\partial\zeta}=\frac{1}{4\pi G_5}\frac{\pi^2T^2}{B^2}e^{4V(r_h)+W(r_h)}.\notag\\
\end{equation}
Using \eqref{kubo_E} and \eqref{kubo_HK}, we obtain $\sigma_\parallel$ and $\sigma_\perp$ in terms of horizon quantities,
\begin{equation}\label{sigma_mp}
\sigma_\parallel=\frac{1}{4\pi G_5}e^{2V(r_h)-W(r_h)},\qquad\sigma_\perp=\frac{1}{4\pi G_5}e^{W(r_h)}.
\end{equation}
We have confirmed that \eqref{sigma_mp} agrees with our numerical results in the hydrodynamic regime for arbitrary $B$.

\end{document}